\documentstyle[11pt,aaspp4]{article}

\newcommand {\be} {\begin{equation}}
\newcommand {\ee} {\end{equation}}

\def\refitem{\par\parskip 0pt\noindent\hangindent 20pt}
\input psfig.sty

\begin{document}

\title{STRUCTURE OF THE CIRCUMNUCLEAR REGION OF SEYFERT 2 
GALAXIES REVEALED BY RXTE HARD X--RAY OBSERVATIONS OF NGC 4945}

\author{
G. Madejski$^{1,2}$, P. \.{Z}ycki$^{3}$, C. Done$^{4}$, 
A. Valinia$^{1,2}$, P. Blanco$^{5}$, R. Rothschild$^{5}$, B. Turek$^{1,6}$}

\affil{
$^1$ Laboratory for High Energy Astrophysics, NASA/GSFC, Greenbelt, MD
20771, USA\\
$^2$ Dept. of Astronomy, University of Maryland, College Park \\
$^3$ Nicolaus Copernicus Astronomical Center, Bartycka 18, 00-716
Warsaw, Poland \\
$^4$ University of Durham, Physics Dept., Durham, DH1 3LE, UK \\
$^5$ Center for Astrophysics and Space Sciences, Univ. of 
California / San Diego, LaJolla, CA\\
$^6$Dept. of Physics, Stanford University, Palo Alto, CA}

\vskip 0.4 cm

\centerline {\sl Submitted to the Astrophysical Journal (Letters)} 

\begin{abstract}

NGC~4945 is one of the brightest Seyfert galaxies on the sky at 100
keV, but is completely absorbed below 10 keV, implying an optical 
depth of the absorber to electron scattering of a few;  its absorption 
column is probably the largest which still allows a direct view of the 
nucleus at hard X--ray energies.  Our observations of it with the Rossi 
X--ray Timing Explorer (RXTE) satellite confirm the large absorption, 
which for a simple phenomenological fit using an absorber with Solar 
abundances implies a column of $4.5^{+0.4}_{-0.4} \times 10^{24}$ cm$^{-2}$.
Using a a more realistic scenario (requiring Monte Carlo modeling of 
the scattering), we infer the optical depth to Thomson scattering 
of $\sim 2.4$.  If such a scattering medium were to subtend a large
solid angle from the nucleus, it should smear out any intrinsic 
hard X--ray variability on time scales shorter than the light travel 
time through it.  The rapid (with a time scale of $\sim$ a day) 
hard X--ray variability of NGC~4945 we observed with the RXTE 
implies that the bulk of the extreme absorption in this object does {\sl
not} originate in a parsec-size, geometrically thick molecular
torus.  Limits on the amount of scattered flux require that the optically
thick material on parsec scales must be rather geometrically thin,
subtending a half-angle $< 10^\circ$.  This is only marginally
consistent with the recent determinations of the obscuring column in 
hard X--rays, where only a quarter of Seyfert~2s have columns which 
are optically thick, and presents a problem in accounting for the 
Cosmic X--ray Background primarily with AGN possessing the geometry as 
that inferred by us.  The small solid angle of the obscuring material, 
together with the black hole mass (of $\sim 1.4 \times 
10^6$ $M_{\odot}$) from megamaser measurements, allows a robust
determination of the source luminosity, which in turn implies that 
the source radiates at $\sim 10$\% of the Eddington limit. 

\end{abstract}


\section{Introduction}

Our current best picture of nuclei of Seyfert galaxies includes 
the central source (i.e. black hole, accretion disk and broad line region)
embedded within an optically thick molecular torus (cf. Antonucci \&
Miller 1985).  The object is classified as a Seyfert~1 for
viewing directions which lie within the opening angle of the torus, 
so that there is a direct view of the nucleus, and as a Seyfert~2 for 
directions intersecting the obscuring material.  The torus absorbs the 
optical, UV and soft X--ray nuclear light, so the nucleus
in Seyfert~2s can only be seen at these energies through scattered
radiation.  One of the central questions still under debate in the
context of the Unification Models of the two types of Seyferts is:
how optically thick is this putative torus?  

NGC~4945 is a nearby (3.7 Mpc;  Mauersberger et al. 1996) edge--on 
galaxy. It has strong starburst activity, producing intense IR 
emission concentrated in a compact nuclear region (Rice et al. 
1988;  Brock et al. 1988), and a ``superwind'' outflowing along 
the minor axis of the galaxy (Heckman, Armus, \& Miley 1990).  
It also has an active nucleus, first seen unambiguously in Ginga 
X--ray observations (Iwasawa et al. 1993), confirming the 
Seyfert~2-type classification.  These data showed a heavily obscured, 
strong hard X--ray source above 10 keV, confirmed by the 
CGRO OSSE observations 
(Done, Madejski, \& Smith 1996) which in turn revealed that NGC~4945 
is one of the brightest extragalactic sources in the sky at 100 keV!  
The absorbing column, a few $\times$ $10^{24}$ cm$^{-2}$, is among the 
largest which still allows a direct view of the nucleus at hard X--ray 
energies.
If such a scattering medium were to subtend a large solid angle from 
the nucleus, it would smear out any intrinsic hard X--ray variability 
on timescales shorter than the light travel time through it.  

The presence of the Seyfert nucleus is further supported by the fact
that the object is a megamaser source (detected in the H$_2$O bands) 
implying an edge--on geometry, but one of the key features which makes 
its study so important is that it is one of only four AGN where the 
black hole mass can be constrained (at $\sim 1.4 \times 10^6$ $M_{\odot}$) from
detailed mapping of the  megamaser spots (Greenhill, Moran, \&
Herrnstein 1997).  As such, it is one of a few unique sources 
where the luminosity in Eddington units can be reliably estimated.  

Below we present the data from the RXTE, confirming the large
absorbing column, but also revealing large amplitude hard X--ray 
variability on a time scale of days.  A distant absorber with an
appreciable optical depth, subtending a large solid angle as seen by
the nucleus, would smear out the rapid variability on time scales
shorter than the light travel time through such an absorber.  Given
our data, we conclude that the optically thick absorber cannot be 
both distant and geometrically thick.  

\section{Observations:  Spectrum and Variability}

NGC~4945 was observed by the Rossi X--ray Timing Explorer (RXTE)
satellite for about a month, starting on October 8, 1997.  
The observations included 38 pointings of $\sim 2000$ s each, 
taken about once per day.  
The Proportional Counter Array (PCA) and 
High Energy X--ray Timing Experiment (HEXTE) data were reduced using 
standard procedures.  For the PCA data, this included an extraction
in the {\tt standard2} mode using the {\tt ftool saextrct};  the 
estimation of the background was done via {\tt ftool pcabackest}, using
the background model ``L7'' (model files {\tt 
pca\_bkgd\_faint240\_e03v03.mdl} and {\tt pca\_bkgd\_faintl7\_e03v03.mdl}).  
For consistency, we use data only from 3 PCA detectors, which were 
active during all pointings.  For HEXTE, the background is collected 
simultaneously by switching two halves of the array on- and off-source 
every 16 seconds;  both source and background files were extracted 
using the {\tt ftool seextrct}, and dead-time corrected using {\tt 
ftool hxtdeadpha}.  The total ``good data'' intervals were:  69,280 s for
PCA, 18,364 s and 19,060 s for HEXTE clusters A and B.  

The summed background-subtracted PCA and HEXTE files were fitted with 
a phenomenological model including a hard power law with low-energy 
photoelectric cutoff (using the cross-sections and abundances as 
given in Morrison \& McCammon 1983) and a high-energy exponential cutoff
(assumed to be at an energy $E_c$ of 100 keV, in agreement with the
high energy Seyfert spectra;  e.g., Zdziarski et al. 1995).  Our model also 
includes a Gaussian Fe K emission line, plus a soft component, which 
we modeled as another power law.  In our fits, we used the PCA data
corresponding to the energy range of 3 to 30 keV, and HEXTE data for
20 to 100 keV.  The resulting fit (cf. Fig.~1) was essentially consistent
with the Ginga / OSSE results of Done et al. (1996).  The hard power
law (with $E_c$ of 100 keV) showed an 
energy index of $0.45 \pm 0.1$, with absorption of $4.5 \pm 0.4 \times 
10^{24}$ cm$^{-2}$, and an observed 10 - 50 keV flux of $1 \times 
10^{-10}$ erg cm$^{-2}$ s$^{-1}$, with resulting $\chi^2 = 80.6 / 75$ 
d.o.f.  The Fe K line energy was at $6.38 \pm 0.05$ keV, with an 
intrinsic width $\sigma$ of $ 0.37 \pm 0.13$ keV, and a flux
of $0.9 \times 10^{-4}$ photons cm$^{-2}$ s$^{-1}$.  Allowing the
cutoff to be unconstrained yielded the best fit of $90^{+130}_{-30}$
keV.  Regarding the soft component, its energy power law index of 
was $0.57 \pm 0.15$, with the 1 keV monochromatic flux of of 0.001 
photons cm$^{-2}$ s$^{-1}$ keV$^{-1}$.  It is important to note that 
the PCA field of view is about 1 deg$^{2}$, so at least a fraction 
of the Fe K line and soft component flux could have arisen from the 
more extended (non-nuclear) region, a likely possibility given the 
starburst nature of the galaxy.  Furthermore, given its modest flux, 
which over the 2 -- 10 keV band is $6 \times 10^{-12}$ erg cm$^{-2}$ 
s$^{-1}$ -- only three times that of the $1 \sigma$ fluctuations of the 
Cosmic X--ray Background on the angular scale of the PCA field of view 
-- we caution that any detailed spectral analysis of the PCA data for
this soft component is unreliable.  Nonetheless, we can clearly reject
the hypothesis that the entire flux of this soft component is due to
some kind of a ``leaky absorber,'' as it does not appear to vary.  

The spectral analysis above confirmed the results of Done et al. (1996)
that the source spectrum consists of the soft, relatively faint, 
unabsorbed component, a bright, heavily absorbed (hard) component, and a
strong Fe K line.  With that, we studied the variability of each
component separately.  The flux of the soft continuum component 
(below 8 keV) is consistent with being constant;  this is also true 
of the Fe K line.  The hard component (8 -- 30 keV), on the other
hand, is highly variable, with a factor of 4 change in 10 days, and a
factor of 1.7 -- 2 in 1 day between the minimum and maximum flux.  
This is plotted in Fig.~2.  This light curve (collected with 3 PCUs,
and binned on 1 day intervals) shows RMS variance ($1 \sigma$) of 0.82 
cts s$^{-1}$.  Unfortunately, the source was too faint to study the 
variability with HEXTE.  

We investigated if the variability could be due to instrumental
effects, and specifically, imprecise background subtraction.  To
assess this, we also analyzed in an analogous manner the hard 
(8 -- 30 keV) light curves binned in 1 day intervals of a cluster 
of galaxies Abell 754 (cf. Valinia et al. 1999) and a faint quasar 
PG1211+143, which shows very little flux in the PCA data above 8 keV 
(Netzer, Madejski, \& Kaspi in prep.).  Analysis of 9 data points 
collected over 9 days for A754 (from which no source variability is 
expected) yields $\sigma = 0.018$ cts s$^{-1}$.  For PG1211+143, we 
had 32 pointings spread nearly uniformly over 6 months.  These data, 
where some intrinsic source variability may be present, yield 
$\sigma = 0.13$ cts s$^{-1}$;  we consider this a conservative upper 
limit to the instrumental effects, and thus deem the rapid hard X--ray
variability of NGC~4945 with $\sigma$ of 0.82 cts s$^{-1}$ highly 
significant.  

Could this variability be due to varying absorption?  We examined 
this possibility by modeling separate spectra from high and low count
rate observations. To improve statistics, we co-added the PCA spectra from 
a number of individual observations with highest and lowest count rates.
The two resulting spectra were then modeled using the Monte Carlo 
absorption model discussed below.  We assumed first that the intrinsic 
source spectrum (and normalization) is the same for both, but 
the absorption is different and, secondly, that the normalization of the
intrinsic source has changed while the absorption stayed constant.
The first hypothesis yielded $\chi^{2} = 401/110$ dof, while the second one
yielded $\chi^{2} = 143/110$ d.o.f. This clearly shows 
that the variability is intrinsic to the unabsorbed nucleus.  

\section{Discussion}

The variability we see in NGC~4945 is then entirely compatible with 
that expected from the {\sl intrinsic} source, with no significant 
scattered delay by a distant material.  If this absorber has an 
appreciable Compton thickness (as is the case here), and if it 
subtends a large solid angle to the X--ray source, then it 
should intercept and scatter a large fraction of the flux.  If it is
also distant from the nucleus, the light travel time effects
will ``wash out'' any intrinsic variability on time scales shorter
than the light travel time through the absorber.  Conversely, 
a structure with much smaller scale height subtends a much smaller 
solid angle, making scattering less important. This also 
precludes the observed X--rays to be purely due to Compton 
reflection, as this would require a contrived geometry of the 
reflector with respect to the primary X--ray source:  the reflector 
would have to be located very close to a completely covered central 
source.  We note that by comparison, a well studied unabsorbed Seyfert~1 
NGC~3516 showed X--ray variability with a similar fractional 
amplitude on $\sim 10 \times$ longer time scales (Nandra 
\& Edelson 1999) than seen in NGC~4945.  While the black hole mass 
in NGC~3516 is not as well known, it is estimated to be 
$\sim 10^7$ $M_{\odot}$.  The fact that the ratio of variability time
scales is roughly the same as the ratio of nuclear masses -- as
expected for accreting black holes -- further supports our conclusion 
that the hard X--ray variability we see in NGC~4945 is intrinsic.  

With the optical depth to electron scattering of a few, the shape of 
the absorption cutoff would be different than expected from pure 
photoelectric absorption, and the detailed shape of the emergent 
spectrum depends on the geometry.  We model this numerically with a 
Monte Carlo code as given in Krolik, Madau, \& \.Zycki (1994), 
assuming a torus with square cross-section where the half-angle 
subtended by the torus, $\theta_0$, its optical depth to electron 
scattering $\tau_{\rm e}$, and the power law index of the incident 
energy spectrum $\alpha$ are free parameters (the Comptonization 
cutoff is set to 100 keV as above).  The results of our fits (using
the PCA data over the range of 3 -- 20 keV, and HEXTE data as above) 
are shown in Table 1, where the 90\% confidence regions on $\Gamma$ 
and $\tau_{\rm e}$ are typically 0.1.  

A small scale height absorber 
($\theta_0 \sim 10^\circ$) gives $\tau_{\rm e} = 2.4$, compared
to a large scale height ($\theta_0 \sim 80^\circ$) which gives 
$\tau_{\rm e} = 2.1$. 
With an iron abundance of twice Solar, these fits change to 
$\tau_{\rm e} = 1.7$, and 
$\tau_{\rm e} = 1.5$, respectively.  (Since the photoelectric 
cutoff present in our data is mainly sensitive to the column density of Fe, 
larger-than-Solar abundance of Fe would make us overestimate 
the true absorbing column if we assumed Solar abundances, and vice-versa.)  
While statistically these might marginally favor 
the large scale height absorber, we consider that all the fits are 
probably equally likely given that modeling the spectrum with a fixed 
cutoff energy may introduce systematic uncertainties.  Our 
calculations include the Fe K emission line produced by the torus but 
we also include an additional Fe line (such as may be expected to 
arise in the photoionized scattering medium).  Those calculations  
also imply that large ($> 4 \times$ Solar) Fe abundances can be
excluded:  they would imply a stronger Fe K line than is seen in our
data. In reality, this limit is probably more stringent, given the fact that at
least a fraction of the Fe K line originates in a more extended
region.  

These Monte Carlo results also give the distribution of the number of
scatterings which the photons undergo before reaching the observer
positioned in the equatorial plane.  
This is key in determining the solid angle subtended by the optically
thick absorber, and thus its vertical size scale.  Fig.~3
shows the fraction of the observed photons that underwent 0, 1, 2, 3, 
etc. scatterings before reaching the observer for eight values of
$\theta_0$ as discussed above (cf. Table~1), 
with the solid and dotted lines showing the results for Solar and
twice Solar abundance of iron, respectively.  The fraction of 
photons which arrive without being scattered is 19\% and 63\% 
respectively for a ``thick'' ($\theta_0 = 80^\circ$) and ``skinny'' 
($\theta_0 = 10^\circ$) torus.  For $2 \times$ Solar abundance of 
Fe these numbers are 32\% and 75\%.  The data in Fig.~2 imply 
that fewer than 40\% of the observed photons are scattered over path 
lengths longer than 1 light day, so the half-angle subtended by the 
optically thick material is less than $\sim 10^\circ$. 
This implies a rather small scale height, and perhaps 
is due to the same material which produces the H$_{2}$O maser emission 
(Greenhill et al. 1997). 

The details of the geometry of Seyfert~2s are important in the
assessment of their contribution to the Cosmic X--ray background, as
the heavily absorbed AGN were postulated to make up the bulk of it 
(cf. Krolik et al. 1994; Madau, Ghisellini, \& Fabian 1994; Comastri 
et al. 1995).  The value of $\theta_0$ of $10^\circ$ is marginally 
consistent with the torus geometry inferred from recent observations 
at hard X--ray energies.  These observations show that Seyfert~2s 
outnumber Seyfert~1s by a factor of 4:1, while more than a quarter 
of Seyfert~2s have a column which is optically thick (see Giommi et 
al. 1998, and Gilli, Risaliti, \& Salvati 1999 and references therein).  
Assuming a single geometry for the Seyfert nuclei where the torus 
has a rectangular cross-section, we can attempt to reproduce these 
ratios by assuming that if the central flux barely ``grazes'' the 
torus, we classify the object as {\sl any} Seyfert~2, while an object 
is an optically thick Seyfert~2 only if the line of sight encounters 
the entire radial distance in the torus.  In this context, requiring 
such 1:4:1 ratio would then imply that the torus is somewhat flattened, 
with outer radius of 6.5 $\times$ its inner radius and equatorial
height of 1.3 $\times$ its inner radius.  In this scenario, all 
Seyfert~2s are then seen at angles smaller than $\sim 50^\circ$ from 
the plane, while the optically thick Seyfert~2s are confined to angles 
of $\le 12^\circ$.  We repeated the Monte--Carlo calculations with
this rectangular geometry and find that the fraction of scattered
photons is $\sim$ 50\%, still too large to match the observed hard
X--ray variability. A further problem arises if
this is indeed a universal geometry for all Seyferts as the
Thomson depth of the absorber is large, $\sim 1.7$ in the 
more restrictive $2 \times$ Solar case.  This absorber reduces the 
unabsorbed flux $\sim$ five-fold or more, and this is not consistent with 
the 1:4:1 ratio, by a large margin.  We thus conclude that a
population of AGN with 
geometry very nearly that of NGC~4945 cannot make up the CXB.  
Instead, significantly larger fraction of the heavily obscured AGN 
is required (implying a large solid angle subtended by the absorber) 
than implied from the rather small $\theta_0$ inferred by us.  One
plausible scenario would have the local optically thick Seyfert~2s 
surrounded by absorbers that already collapsed to an accretion 
disk, while in the more distant objects -- in the earlier stages of 
evolution -- such absorbers had larger vertical extent.  

Alternately, the absorption could come from a structure which is
much closer to the nucleus.  The variability limits impose constraints
on the amount of scattered flux that is lagged on time scales of more
than 1 day, and thus they do not constrain the height of any structure which 
is $<< 1$ light day from the X--ray source (corresponding to a 
distance of $<< 10^4$ Schwarzschild radii for the mass of the black 
hole of $\sim 10^6$ $M_{\odot}$). While a very geometrically thick 
accretion disk cannot be ruled out from our data, there are 
considerable theoretical difficulties in maintaining a structure 
with a large height scale. It is far easier to envisage a structure 
with a small height scale such an accretion disk with outer
regions somewhat thickened by instabilities resulting from radiation
pressure warping (cf. Maloney, Begelman, \& Pringle 1996).  

With these arguments for the Thomson-thick absorber subtending a small 
solid angle in NGC~4945, we can now estimate the true luminosity of 
the source.  The Monte Carlo simulations show that the intrinsic 
flux must be substantially larger than that observed (by a factor 
of $e^{\tau_{\rm e}}$, or $\times 11$ for Solar abundances) because 
any scattered nuclear photons are lost from the line of sight.  
The 1-500 keV flux, corrected for photoelectric absorption alone, 
is $\sim 5 \times 10^{-10}$ erg cm$^{-2}$ s$^{-1}$, so correcting for 
the Thomson opacity yields a 1-500 keV intrinsic X--ray 
luminosity of $10^{43}$ erg~s$^{-1}$.  Assuming that the thermal (opt/UV/EUV) 
emission from the accretion disk is roughly equal to the hard X--ray emission
gives a total bolometric luminosity of the nucleus of $\sim 2 \times 
10^{43}$ erg~s$^{-1}$.  With this, and $M_{\rm BH}$ of $1.4 \times
10^6$ $M_{\odot}$, the source is radiating at $\sim$ 10\% of the 
Eddington luminosity;  even if the abundances are twice-Solar (which 
yields $\tau_{\rm e}$ of 1.7), $L/L_{\rm Edd}$ is at least $\sim$ 5\%.  
(We note that similar $L/L_{\rm Edd}$ was also inferred by 
Greenhill et al. (1997), but the discovery of the rapid hard X--ray 
variability allows us to determine the source luminosity more
accurately, as now we know that relatively few photons are scattered 
{\sl back} to the line of sight.)  NGC~4945 is one of the few AGN where this 
quantity can be calculated robustly, since the mass of the central object is
{\sl known}, although we are aware that the value of its central mass is not 
as accurate as for the famous megamaser NGC~4258;  the resulting 
uncertainty in the estimate of $L/L_{\rm Edd}$ may be a factor of 2, 
comparable to the effects of the unknown Fe abundance or the ratio of 
$L_{\rm Tot} /L_{\rm X-ray}$.  The resulting $L/L_{\rm Edd}$ is comparable 
to that inferred for the well studied Seyfert~2 NGC~1068, although since the
absorber is completely opaque even to hard X--rays, the central luminosity can
be estimated only indirectly.  These two AGN are at the opposite end of 
the scale to NGC~4258, which radiates at $\sim 10^{-4}$ $L_{\rm Edd}$ 
or less (Lasota et al.\ 1996).  The mass accretion rates inferred 
from those values of $L/L_{\rm Edd}$ put strong constraints on
possible underlying radiation mechanisms.  While the recently popular 
advective disk models can produce X--ray hot flows up to about 10\% 
of the Eddington limit, these collapse at higher $L/L_{\rm Edd}$:  
our data show that NGC~4945 lies perilously close to this limits.  

\bigskip

{\bf ACKNOWLEDGEMENTS:} We thank the
RXTE satellite team for scheduling the observations allowing the daily
sampling, Tess Jaffe for her help with the RXTE data reduction via
her indispensable script {\tt rex}, and Dr. Julian Krolik for his
helpful comments on the manuscript.  This project was partially supported by 
ITP/NSF grant PHY94-07194, NASA grants and contracts to University of
Maryland and USRA, and the Polish KBN grant 2P03D01816.

\setcounter{table}{0}
\begin{deluxetable}{ccccccccc}
\tablewidth{0pc}
\tablecolumns{9}
\tablecaption {Results of Monte Carlo Fits to the RXTE Data for NGC~4945 
for Assumed Solar and $2 \times $ Solar Fe Abundances}
\tablehead{
\colhead {Assumed torus} & \multicolumn {2} {c} {Fitted spectral} & 
\multicolumn {2} {c} {Fitted optical} & \multicolumn {2} {c} { $\chi^2$} & \multicolumn {2} {c} {Fraction of 
detected} \\
\colhead {half-angle $\theta_0$} & 
\multicolumn {2} {c} {index $\alpha$} & 
\multicolumn {2} {c} {depth $\tau_{\rm e}$} & \multicolumn {2} {c} {76 d.o.f.} & 
\multicolumn {2} {c} {{\sl unscattered} photons }  \\
\colhead {(degrees)} & 
\colhead {A$_{\rm Fe} = 1$} & \colhead {A$_{\rm Fe} = 2$} & 
\colhead {A$_{\rm Fe} = 1$} & \colhead {A$_{\rm Fe} = 2$} & 
\colhead {A$_{\rm Fe} = 1$} & \colhead {A$_{\rm Fe} = 2$} & 
\colhead {A$_{\rm Fe} = 1$} & \colhead {A$_{\rm Fe} = 2$}  
}
\startdata
   80  &  0.7  &  0.75 &  2.1  & 1.5 &    68.5 & 68.5 &  19\% & 31\% \nl
   70  &  0.7  &  0.75 &  2.1  & 1.5 &    69.8 & 68.3 &  22\% & 35\% \nl
   60  &  0.7  &  0.75 &  2.2  & 1.5 &    71.5 & 72.9 &  24\% & 39\% \nl
   50  &  0.8  &  0.75 &  2.2  & 1.6 &    70.9 & 69.6 &  28\% & 42\% \nl
   40  &  0.7  &  0.75 &  2.2  & 1.6 &    70.9 & 72.7 &  33\% & 48\% \nl
   30  &  0.8  &  0.75 &  2.3  & 1.6 &    76.6 & 74.6 &  38\% & 55\% \nl
   20  &  0.8  &  0.75 &  2.4  & 1.6 &    74.5 & 77.0 &  47\% & 63\% \nl
   10  &  0.8  &  0.8  &  2.4  & 1.7 &    75.4 & 79.4 &  63\% & 76\% \nl
\enddata
\end{deluxetable}

\vskip 1 cm

\centerline {\bf Figure Captions}

\refitem{\bf Fig.~1:} 
Broad-band unfolded X--ray spectrum of NGC~4945 as measured with the 
RXTE PCA and HEXTE instruments.  The data were fitted with a 
phenomenological model which includes a hard power law component
photo--electrically absorbed by neutral gas with Solar abundances at a
column of $4.5 \pm 0.3 \times 10^{24}$ cm$^{-2}$, 
with photon spectral index $\Gamma = 1.45^{+0.1}_{-0.1}$,
exponentially cutting off at 100 keV, plus a non-variable 
soft component (assumed to be a power law), 
and a Fe K line.  The observed 8 -- 30 keV flux of the hard component
is $5 \times 10^{-11}$ erg cm$^{-2}$ s$^{-1}$.  

\vskip 1 cm

\refitem{\bf Fig.~2:} Hard X--ray light curve of the Seyfert~2 galaxy 
NGC~4945 measured with the RXTE PCA instrument, showing a rapid,
large amplitude flux variability.  Plotted are data from all three
layers of three PCA detectors that were turned on during all 
pointings, over the energy channels nominally corresponding to the range 
of 8 -- 30 keV.  

\vskip 1 cm

\refitem{\bf Fig.~3:} Fraction of the observed photons reaching an
observer located in the equatorial plane of a torus plotted against 
the number of scatterings that those photons encountered before  
reaching an observer.  The angle given in each 
panel is the vertical half-angle $\theta_{\rm 0}$ subtended by the torus
as seen from the central source.   Iron abundance (relative to Solar)
is assumed to be 1 (solid line) and 2 (dotted line), with 
$\tau_{\rm e}$ equal to the best fit value for a given Fe abundance
and $\theta_0$ (cf. Table 1).
Since the fractional amplitude of variability on short time scales 
(cf. Fig.~2) is large ($>$ 60\%), $\theta_{\rm 0}$ of the optically
thick 
structure must be small, so that majority of the photons reaching an observer
are not scattered.

\vfill\eject
\centerline{\psfig{file=madejski_fig1.ps,height=5.3 in,angle=270}}
\vfill\eject
\centerline{\psfig{file=madejski_fig2.ps,height=5.3 in,angle=270}}
\vfill\eject
\centerline{\psfig{file=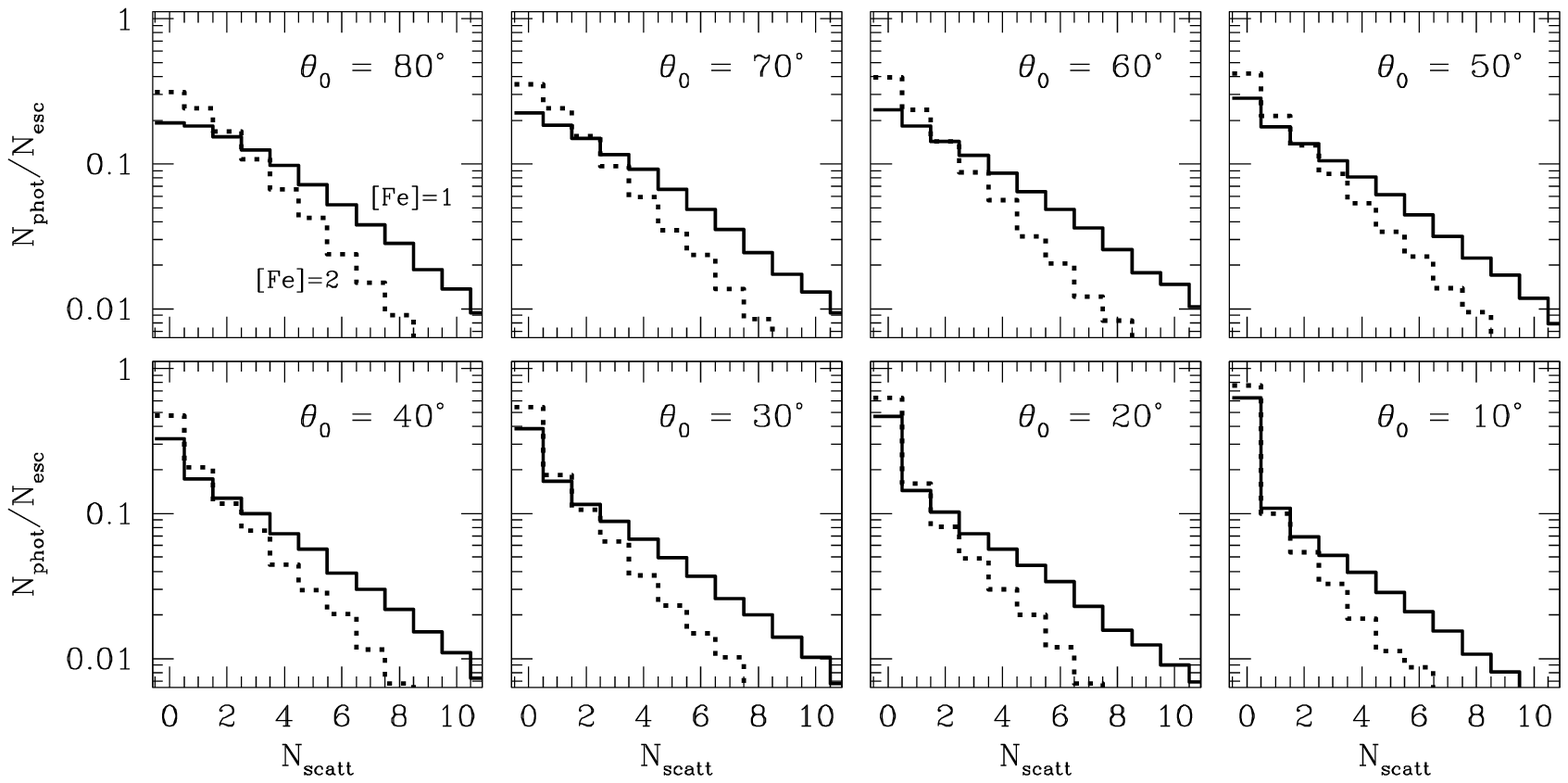,height=8 in,angle=0}}

\enddocument